\documentclass[11pt]{article}

\usepackage{amsmath}
\usepackage{amssymb}
\usepackage{lineno}

\usepackage{graphicx}

\usepackage[square, numbers]{natbib}
\usepackage{color, cancel}

\usepackage{setspace}
\renewcommand{\arraystretch}{0.4}
\usepackage[labelfont=bf,labelsep=period,justification=raggedright]{caption}

\makeatletter
\renewcommand{\@biblabel}[1]{\quad#1.}
\makeatother

\date{}

\begin{document}
\bibliographystyle{plainnat}

\begin{flushleft}
{\Large
\textbf{A unifying framework for quantifying the nature of animal interactions}
}
\\
{\footnotesize
Jonathan R. Potts$^{1,a}$,
Karl Mokross$^{2,3,b}$,
Mark A. Lewis$^{1,4,c}$.
\\
\bf{1} Centre for Mathematical Biology, Department of Mathematical and Statistical Sciences, University of Alberta, Canada
\\
\bf{2} {School of Renewable Natural Resources, Louisiana State University Agricultural Center, Baton Rouge, Louisiana, 70803}
\\
\bf{3} {Projeto Din\^amica Biol\'ogica de Fragmentos Florestais. INPA. Av. Andr\'e Ara\'ujo 2936. Petr\'opolis. Manaus. Brazil. 69083-000}
\\
\bf{4} Department of Biological Sciences, University of Alberta, Edmonton, Canada
\\
\bf{a} E-mail: jrpotts@ualberta.ca.  

\bf{b} E-mail: kmokro1@tigers.lsu.edu.

\bf{c} E-mail: mark.lewis@ualberta.ca.
}
\end{flushleft}


\vspace{5mm}\noindent\textbf{Short title:} Unifying animal interaction models

\vspace{3mm}\noindent\textbf{Keywords:} animal movement, collective behaviour, insectivore birds, resource selection, space use, step selection, territoriality, theoretical ecology.








\newpage
\section*{Abstract}

Collective phenomena, whereby agent-agent interactions determine spatial patterns, are ubiquitous in the animal kingdom.  On the other hand, movement and space use are also greatly influenced by the interactions between animals and their environment.  Despite both types of interaction fundamentally influencing animal behaviour, there has hitherto been no unifying framework for the models proposed in both areas.  Here, we construct a general method for inferring population-level spatial patterns from underlying individual movement and interaction processes, a key ingredient in building a statistical mechanics for ecological systems.  We show that resource selection functions, as well as several examples of collective motion models, arise as special cases of our framework, thus bringing together resource selection analysis and collective animal behaviour into a single theory.  In particular, we focus on combining the various mechanistic models of territorial interactions in the literature with step selection functions, by incorporate interactions into the step selection framework and demonstrating how to derive territorial patterns from the resulting models.  We demonstrate the efficacy of our model by application to a population of insectivore birds in the Amazon rainforest.

\newpage
\section*{Introduction}

Recent years have seen an explosion in the number of studies devoted to collective animal movement modelling, largely enabled by the availability of fast computational power and vastly improved tracking data \citep{sumpter2006,sumpter2010}.  They have succeeded in explaining a wide variety of patterns observed in nature due to the movements and interactions of animals \citep{deneubourggoss1989,couzinetal2002}, such as bird flocking \citep{nagyetal2010}, ant raids \citep{deneubourgetal1989}, and fish schooling \citep{hoareetal2004}.  Furthermore, in the last few years, the collective behaviour paradigm has been extended to include territorial patterns, which arise from conspecific avoidance mechanisms rather than those of alignment or attraction \citep{moorcroftlewis,GPH1,PHG3}.

Despite these myriad advancements, animal-interaction models remain disparate and varied, with no formulation of a unifying framework encompassing the variety of interaction mechanisms, direct or mediated, attractive or repulsive.  This makes it difficult to compare models quantitatively and so determine which behavioural aspects are necessary for causing the observed behaviour.  Though several techniques have recently been proposed for selecting between models of collective movement and interactions \citep{sumpteretal2012}, they tend to be system-specific.  For example, fish repulsion-alignment-attraction mechanisms have been measured using several different techniques \citep{herbertreadetal2011,katzetal2011,gautraisetal2012}, as have the geometric nature of their interactions \citep{strandburgpeshkinetal2013} and their decision processes \citep{perezescuderopolavieja2011, mannetal2014}.  Others examples include alignment and leadership decisions in bird flocks \citep{nagyetal2010,ballerinietal2008, pettitetal2013}.  There are also a few theoretical studies aimed at more general application, e.g. \citep{mann2011}.  However, it is not clear whether they can readily be applied to behaviours beyond grouping phenomena such as swarming, flocking or schooling. 

On the other hand, mechanistic models of territorial interactions have typically been analysed by fitting the emergent spatial patterns, rather than the underlying movement processes, to positional data, e.g. \citep{PHG3,moorcroftetal1999}.  Whilst this is a reasonable way of testing hypotheses about the underlying causes of spatial patterns \citep{moorcroftetal2006}, it is not sufficient for concrete quantification of the underlying movement and interaction processes, since many different model processes can give rise to the same emergent spatial patterns.  Furthermore, territorial modelling approaches have hitherto followed two separate paradigms.  The first involves constructing partial differential equations (PDEs) either from details of the underlying movement and interaction processes or from more phenomenological descriptions, and then using these equations to derive territorial patterns mathematically \citep{moorcroftlewis, lewismurray1993, moorcroftetal1999, moorcroftetal2006}.  The second approach is based more on statistical physics, analysing the individual movement and interaction processes themselves in discrete space, without taking a mean-field continuum limit \citep{GPH1,PHG2}.  A recent review explains the biological lessons that can be learned from these models \citep{PL1}.  These  approaches would benefit from unification both with each other and with the rest of the collective behaviour literature.

Parallel to the collective animal literature, many studies have sought to understand and predict space use patterns by examining interactions between animals and their environment.  Resource selection analysis, positing that animal space use is a function of the distribution of underlying resources, is perhaps the widest used tool in this regard, and has a long history \citep{manlyetal2002}.  Recently, this has been integrated with animal movement processes by constructing step selection functions \citep{fortinetal2005,rhodesetal2005,foresteretal2009}, where the selection of resources is constrained by the ability of an animal to move.  Such functions are built by rigorously deriving parameter values from the data using well-developed statistical techniques \citep{thurfjelletal2014}.
Step selection functions, in turn, have been used to build mechanistic models to derive space use patterns from the underlying movement processes and animal-environment interactions \citep{moorcroftbarnett2008}, representing the first step in unifying resource selection with mechanistic models. 

Some studies in the step selection function literature have factored into their analysis either positions of other individuals \citep{vanakaetal2013} or traces left in the environment by animals \citep{latombeetal2013}.  However, to model simultaneously more than one interacting group of animals, so that it is possible to build a mechanistic model to predict the resulting space use patterns, would require having different interacting movement kernels for each group.  For example, in a territorial system there would be one function for each group maintaining a territory.  These would then have to be coupled together so that each function depends on the animals modelled by the other.  

In this paper, we present a modelling framework that unifies movement with both animal-environment and inter-animal interactions.  The inter-animal interactions may either be direct, or mediated by a stigmergent process \citep{theraulazbonabeau1999,GPRL} such as pheromone deposition or visual cues.  Our framework includes as special cases both step selection functions and the two approaches to mechanistic territorial modelling mentioned above.  Though we focus specifically on combining territorial interactions into the concept of step selection, our framework also happens to incorporate a variety of other collective motion models, suggesting far broader application (see Table \ref{cssf_lit_table}).  As such, our framework provides a useful way to codify movement and interaction processes, giving a generic starting point for modelling these processes and a clear way of testing which combinations of them best describe the underlying data.  This will both help future researchers in model construction and provide a concrete means by which to compare and contrast different modelling approaches.

\begin{figure}[ht!]
\centering
\includegraphics[width=80mm]{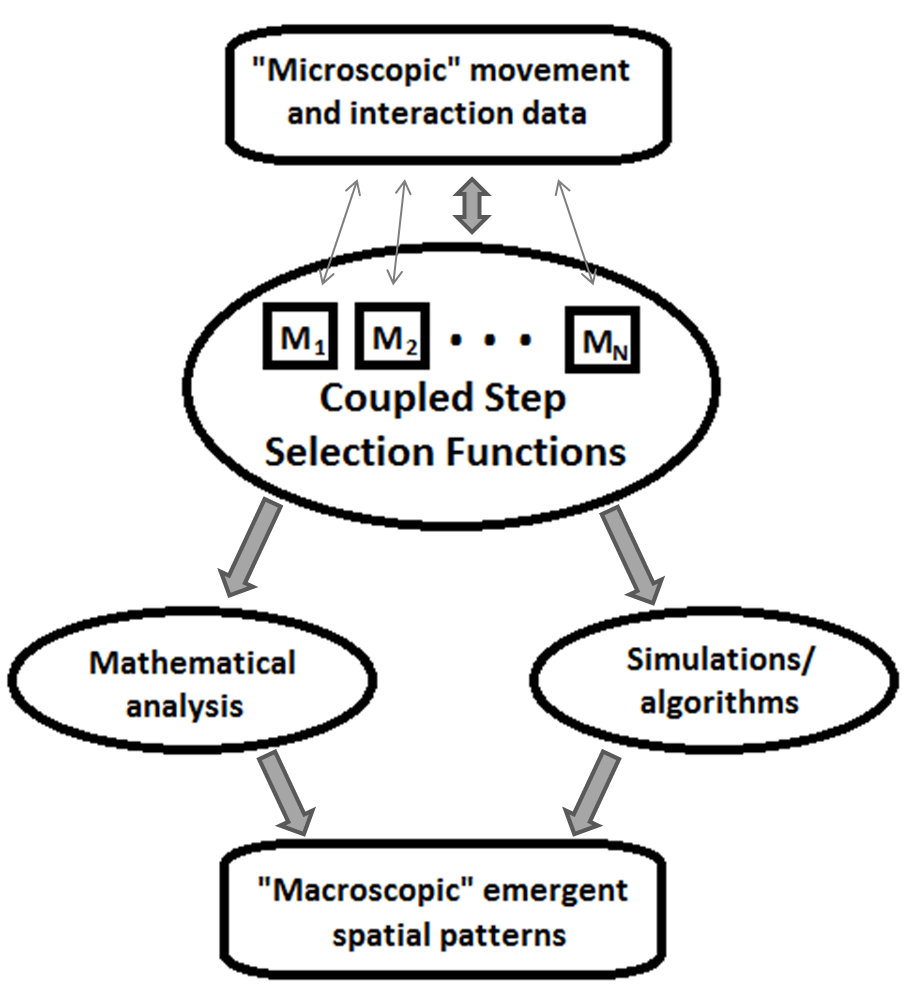}
\caption{{\bf The role of coupled step selection functions in linking movement to emergent spatial patterns.} Different candidate models $M_1,\dots,M_N$ can be tested against `microscopic' movement and interaction data using the techniques in the Methods section.  The best models can then either be simulated or mathematically analysed to derive spatial patterns.  These, in turn, can be compared to the `macroscopic' spatial distributions in the data (see Methods) to test whether the mechanisms being modelled are sufficient for accurate predictions of spatial patterns.}
\label{schematic}
\end{figure}

We show how to use our model to test hypotheses about the interaction mechanisms underlying territorial behaviour, by application to movement data on a community of territorial insectivore bird flocks in the Amazon rainforest.  Parameter values for a model of movement and territorial interactions naturally arise from this hypothesis testing.  This model can then be analysed either using PDE techniques \citep{moorcroftlewis} or by simulation analysis \citep{GPH1,PHG2}.  This enables the spatial territorial patterns to be derived from the underlying movement and interaction processes, which can be compared with spatial data.  We demonstrate how to make this comparison quantitative, thereby giving a technique for determining which the processes are the key drivers of space use in the study population.  As such, the framework used here provides a vital bridge between the selection of models on the individual level and the validation of their emergent features on the population level.  In Fig. \ref{schematic}, we give a schematic that represents the central place of coupled step selection functions for the program of constructing a `statistical mechanics for ecological systems' \citep{levin2012}.

\section*{Methods}

\subsection*{Modeling framework}

\begin{figure}[ht!]
\centering
\includegraphics[width=150mm]{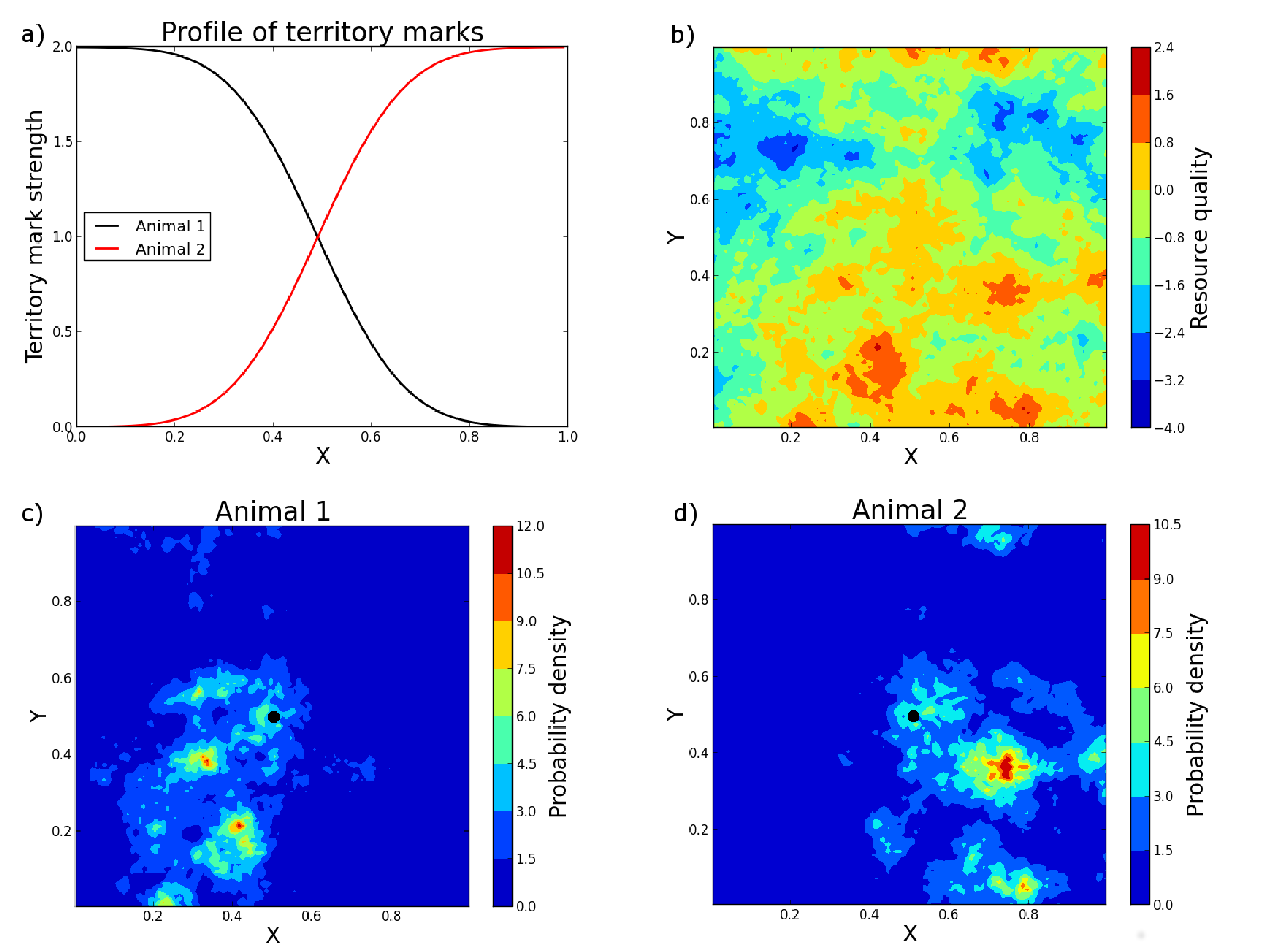}
\caption{{\bf Where next?} A typical coupled step selection function, giving the probability of an animal's next move, dependent on territorial marks and resource quality.  This is determined both by the strength of territorial marks of conspecifics, given in panel (a), and the quality of the resources (b).  The strength of territory marks in this example does not change in the $Y$-direction, so that animal 1 has territory on the left and animal 2 on the right.  The probability of animal 1's (resp. animal 2's) next position after some time interval $\tau$, given that it's current position is in the middle of the landscape (black dot), is shown in panel (c) (resp, panel d).  As each animals moves, it marks the terrain causing the territorial profile to change over time, which in turn influences the other animal's movements.  This causes a coupling between their respective step selection functions.}
\label{hypothetical_landscape}
\end{figure}

{\noindent}Our model is based around the notion of a step selection function \citep{fortinetal2005}.  However, simultaneous modelling of various interacting animals, or groups of animals, requires having a different step selection function for each animal or group.  Therefore, instead of having one function that models all agents, as with previous approaches, we construct a different function for each agent and link them together with a coupling term.  We use the term `agent' here to refer to either a single animal, or a group of animals that are modelled as moving together as a single entity, for example a pack or a flock.  The result is what we call a system of {\it coupled step selection functions} (CSSFs), where each function has the following form
\begin{align}
\underbrace{f_i^{t,\tau}({\bf x}|{\bf y},\theta_0)}_{\text{\begin{tabular}{c}movement \\ probability\end{tabular}}} \propto
\underbrace{\phi_i({\bf x}|{\bf y},\theta_0)}_{\text{\begin{tabular}{c}step length \\ and turning \\ angle\end{tabular}}}
\underbrace{{\mathcal W}_i({\bf x},{\bf y},{\mathcal E})}_{\text{\begin{tabular}{c}environment \\ interactions\end{tabular}}}\underbrace{{\mathcal C}_i({\bf x},{\bf y},{\mathcal P}_i^t)}_{\text{\begin{tabular}{c} collective \\ interactions\end{tabular}}},
\label{cssf_general}
\end{align}
represented pictorially in Fig. \ref{hypothetical_landscape}.  The function $f_i^{t,\tau}({\bf x}|{\bf y},\theta_0)$ is the probability of agent $i$ moving to position ${\bf x}$ at time $t+\tau$, given that the agent was at position ${\bf y}$ at time $t$ and had arrived there on a bearing $\theta_0$.  The term $\phi_i({\bf x}|{\bf y},\theta_0)$ represents the movement process of agent $i$, disregarding the effect of the environment or other agents.  For example, this could contain the step length and turning angle distribution for a correlated random walk \citep{bovetbenhamou1988}.

The function ${\mathcal W}_i({\bf x},{\bf y},{\mathcal E})$ is a weighting function containing information about the desirability of moving across the environment ${\mathcal E}$ from position ${\bf y}$ to ${\bf x}$.  For example, if there is a partial barrier to movement between ${\bf y}$ and ${\bf x}$ then ${\mathcal W}_i({\bf x},{\bf y},{\mathcal E})$ may be lower than if the barrier were not there.  On the other hand, if ${\bf x}$ were in a very desirable habitat for the agent compared to ${\bf y}$ then ${\mathcal W}_i({\bf x},{\bf y},{\mathcal E})$ would be higher than if the habitats were equal in quality.  See \citep{fortinetal2005} for a good example of the variety of animal-environment interactions that can be modelled this way.

The collective aspects of motion, i.e. the agent-agent interactions, are represented by  ${\mathcal C}_i({\bf x},{\bf y},{\mathcal P}_i^t)$.  The term ${\mathcal P}_i^t$ represents both the population positions and any traces of their past positions left either in the environment or in the memory of agent $i$.  For example, if the agents were schooling fish then perhaps the pertinent interactions would be direct \citep{ioannouetal2012}.  However if the agents were ants then ${\mathcal P}_i^t$ might represent the pheromones left by other ants, to which ant $i$ responds by tending to move up the pheromone gradient \citep{deneubourgetal1989}.  As a third example, if the agents were territorial bird flocks then ${\mathcal P}_i^t$ might include the memory that the birds in flock $i$ have of past territorial conflicts or vocalisations.  In most realistic cases, including the ones detailed here, this enables us to convert ostensibly non-Markovian processes, such as memory and correlations, into one-step Markov processes, possibly requiring high dimensions to encapsulate ${\mathcal P}_i^t$ appropriately.

Since $f_i^{t,\tau}({\bf x}|{\bf y},\theta_0)$ is a probability, it must integrate or sum to 1, depending on whether continuous space or discrete space is being used, respectively.  Therefore we use the $\propto$ sign in equation \ref{cssf_general}, noting that this becomes an equality if the right hand side is divided by the integral (continuous space) or sum (discrete space) over the possible target positions ${\bf x}$.

We demonstrate the generality of our formalism by showing that it reduces to ordinary step selection functions \citep{fortinetal2005}, resource selection functions \citep{boyceetal2002}, and a variety of previously published examples of collective motion models.  The latter include models of trail-following ants \citep{deneubourgetal1989}, collective patterns in animal populations through alignment and attraction \citep{couzinetal2002, eftimieetal2007}, and territorial canids \citep{moorcroftlewis,PHG3,PHG2}.

It is possible to generalise equation (\ref{cssf_general}) further by writing the right-hand side as an arbitrary function of ${\bf x}$, ${\bf y}$, $t$, $\theta_0$, ${\mathcal E}$ and ${\mathcal P}_i^t$.  This would enable the construction of dependencies between the three aspects of movement, environmental interactions, and collective interactions.  For example, this could describe the animal's speed varying over time due to seasonal changes, or the turning angle distribution being effected by habitat type, and so forth.  However, the models from both previous collective animal behaviour studies and the step/resource-selection literature tend not to incorporate such dependencies, since they can be written in the form of equation (\ref{cssf_general}).  Therefore, for simplicity, we treat the functions $\phi_i$, $W_i$ and $C_i$ as independent.

\subsection*{Application to bird data}

As a demonstration of how to apply our model, we use movement data on a community of territorial insectivore bird flocks in the Amazon rainforest.  These flocks are multi-species, with around 5-10 mating pairs consistently present sharing a territory \citep{munnterborgh1979}.  Each pair will defend its territory from conspecifics, using a mixture of vocalisations and direct territorial conflicts \citep{jullienthiollay1998}.  The birds from each flock meet together at a `gathering point' at dawn every day, usually in a central position within their territory, from where they forage within the territory for around 11-12 hours, moving together as a flock.

We use flock movement data from eleven different territories to test hypotheses about the territorial interaction mechanisms used by the birds.  We focus, for simplicity, on the vocal aspect of interactions.  Vocalisations make neighbouring flocks aware of areas they have recently visited, causing the neighbours to alter their movement processes in or near these areas.  We test three hypotheses: whether (1) flocks are likely to avoid areas that neighbours have visited in the past, due to the vocalisations made there, (2) flocks tend to move back towards their gathering site having visited such an area, (3) the time since the area was visited by a neighbour affects the response of the flock, so that old vocalisations are ignored.  This demonstrates the ability of our modelling framework to select between competing theories about the nature of interaction mechanisms.

We analysed movement of 11 different flocks in the Amazon rainforest over 3 years during the dry season between June and November.  The study site is about 70 km north of Manaus, Brazil.  They were each tracked for between 4 and 18 days.  The flock positions were recorded every minute during the time that they were active.  Flock activity is conspicuous, so that birds can be followed on foot.  As flocks moved, geolocations were recorded with a hand-held GPS unit (Garmin Vista HCX). The observer maintained a distance of 10-20m from the flocks to ensure no alarm or avoidance behaviour was induced in the birds.

To examine which territorial interaction processes best fit these data, we constructed a coupled step selection function (Eq. \ref{cssf_general}) where the terms $\phi_i({\mathbf x}|{\mathbf y},\theta_0)$ and ${\mathcal W}_i({\mathbf x},{\mathbf y},{\mathcal E})$ were obtained from a previous study on the same population \citep{PMSL1}.  In that paper, we found that setting $\phi_i$ to be a product of the exponentiated Weibull distribution \citep{nassareissa2003} for the step lengths and a von Mises distribution \citep{marshjones1988} for the turning angles fitted the data well.  This led to the following step length and turning angle distribution
\begin{align}
\phi_i({\mathbf x}|{\mathbf y},\theta_0) =& \underbrace{\frac{ac}{b}\left(\frac{|{\mathbf x}-{\mathbf y}|}{b}\right)^{a-1}\exp\left[-\left(\frac{|{\mathbf x}-{\mathbf y}|}{b}\right)^a\right]\left\{1-\exp\left[-\left(\frac{|{\mathbf x}-{\mathbf y}|}{b}\right)^a\right]\right\}^{c-1}}_{\text{step length distribution}} \times \nonumber \\
&\underbrace{\frac{\exp[k\cos(\theta-\theta_0)]}{2\pi\mbox{I}_0(k)}}_{\text{turning angle distribution}},
\label{bird_phi}
\end{align}
where each agent $i$ is an individual flock, $\theta$ is the bearing from ${\mathbf y}$ to ${\mathbf x}$, $a=1.06\pm 0.03$, $b=6.90\pm 0.34$, $c=1.82\pm 0.11$, $k=0.336\pm0.015$ (error bars are 1 standard deviation) and $\mbox{I}_0(k)$ is a modified Bessel function of the first kind.  The best fit model from \citep{PMSL1} for the ${\mathcal W}_i$ term is ${\mathcal W}_i({\mathbf x},{\mathbf y},{\mathcal E})=C({\mathbf x})^\alpha T({\mathbf x})^{-\beta}$, where $C({\mathbf x})$ and $T({\mathbf x})$ are, respectively, the forest canopy height and topography in meters, at position ${\mathbf x}$.  The time-interval $\tau$ is 1 minute and the best fit values for the parameters are $\alpha=0.0952\pm0.037$ and $\beta=1.658\pm0.345$ (error bars are 1 standard deviation).  These were derived by performing the model fit whilst neglecting interaction mechanisms (see \citep{PMSL1} for details).

For the interaction term ${\mathcal C}_i({\mathbf x},{\mathbf y},{\mathcal P}_i^t)$, we set ${\mathcal P}_i^t({\mathbf x})=T_\ast$ if any flock $j\neq i$ is at position ${\mathbf x}$ at time $t$, and ${\mathcal P}_i^t({\mathbf x})=\min\{{\mathcal P}_i^{t-\tau}({\mathbf x})-\tau,0\}$ otherwise.  Here, $T_\ast$ represents the amount of time a bird will remember a conspecific bird call from a particular location, and so respond to this memory when in that location.  The Cinerous Antshrike from each flock tends to make a call about every 2-5 minutes, which can be detected by other birds at a distance of about 50 meters (Karl Mokross, pers. obs.).  In our model, we implicitly assume, for simplicity, that birds make calls each time they move and that they are always heard by neighbouring flocks.  Notice that this construction is similar in mathematical form to the territoriality model from \citep{PHG3}, used to uncover behavioural mechanisms in a red fox ({\it Vulpes vulpes}) population.  However, in that study, $T_\ast$ represented the longevity of scent cues rather than the memory of vocalisations.



To test hypothesis (1), we examined whether using the following coupling function
\begin{align}
{\mathcal C}_i({\mathbf x},{\mathbf y},{\mathcal P}_i^t)=\{[T_\ast-{\mathcal P}_i^t({\mathbf x})]/T_\ast\}^\gamma
\label{ci_hyp1}
\end{align}
gives a better fit to the data than the case of no interactions, ${\mathcal C}_i({\mathbf x},{\mathbf y},{\mathcal P}_i^t)=1$.  For hypothesis (2), we used the following coupling function
\begin{align}
{\mathcal C}_i({\mathbf x},{\mathbf y},{\mathcal P}_i^t)=V(\underbrace{\kappa I[{\mathcal P}_i^t({\mathbf y})>0]}_{\text{\begin{tabular}{c}attractive \\ strength\end{tabular}}},\underbrace{\theta-\theta_g}_{\text{\begin{tabular}{c}direction \\ bias\end{tabular}}})
\label{ci_hyp2}
\end{align}
with $T_\ast=\infty$, where $V(\lambda,\psi)$ is a von Mises distribution (a single-mode distribution often used in the ecology literature for biasing angles \citep{marshjones1988}), $I[X]$ an indicator function equalling 1 if $X$ is true and 0 otherwise, $\theta$ is the bearing from ${\mathbf y}$ to ${\mathbf x}$ and $\theta_g$ is the bearing from ${\mathbf y}$ to the gathering point.  For hypothesis (3), we used the coupling function from Eq (\ref{ci_hyp2}), but with $T_\ast$ a finite free parameter, to test whether allowing $T_\ast$ to be finite significantly improves the fit.

\renewcommand{\arraystretch}{1}

We fitted the various models to the data using a maximum likelihood technique, whereby we found the free parameters that maximise the product over $i$ and $n$ of $f_i^{t_{i,n},\tau}({\mathbf x_{i,n+1}}|{\mathbf x_{i,n}},\theta_{i,n})$, where ${\mathbf x}_{i,0},\dots,{\mathbf x}_{i,N_i}$ are the positions of flock $i$ at times $t_{i,0},\dots,t_{i,N_i}$.  To find this maximum, we used the Nelder-Mead simplex algorithm as implemented in the Python \texttt{maximize()} function from the SciPy library \citep{scipy}.  For hypothesis (1), the free parameters are $T_\ast$ and $\gamma$.  For hypothesis (2), the free parameter is $\kappa$, and for (3) they are $T_\ast$ and $\kappa$.  The $p$-values for hypothesis testing were obtained using the likelihood ratio test.

One of the strengths of the coupled step selection function approach is that the result of hypothesis testing and/or model selection naturally gives rise to a mechanistic movement model, given by the particular version of equation (\ref{cssf_general}) that corresponds to the best fit model and parameter values.  This enables one to determine the space use (i.e. home range) patterns that emerge from the model.  We test whether the patterns that emerge from the best model that includes resource selection, topographical selection and territorial interactions are a significantly better fit to the data than the same model without the territorial interactions.

To do this, we constructed a simulation model for the bird flocks, whose movements each step are determined by drawing from the time-dependent probability distribution from Eq. \ref{cssf_general} with the best-fit parameter values found by the hypothesis testing technique above.  Since each flock gathers in one particular place each day, and moves around the terrain for a total of about eleven-and-a-half hours during the day, we started the simulated birds at the gathering point and ran the simulation for 690 time steps, each step representing $\tau=1\mbox{ minutes}$ (giving 11 hours 30 minutes in total), taking a note of all the positions at which the flock landed after each step.  We repeated this 100 times, representing 100 days, giving 69,000 simulated positions for each flock, from which we calculated home ranges using the Kernel Density Estimation (KDE) method.  We also ran identical simulations except where the model has ${\mathcal C}_i({\mathbf x},{\mathbf y},{\mathcal P}_i^t)=1$, so that no territorial interactions were included.

To test which model performed better at predicting space use, we compared the Kullback-Leibler (K-L) distance \citep{burnhamanderson} between each model's KDE distribution and the KDE distribution for the data.  The K-L distance differs by a constant from $1/2$ times the average Akaike Information Criterion (AIC) of a single sample from the data's KDE distribution (see \citep{burnhamanderson} for details).  Therefore the difference in AIC ($\Delta$AIC) for two different models of the same data distribution can be thought of as twice the difference in K-L distance, by considering a single KDE distribution as a single data sample.  We have 11 flocks, so 11 KDE distributions.  The $\Delta$AIC is twice the sum of the differences in K-L distance across these flocks.  We use this value to assess whether the resulting model is better at predicting {\it space use}, as opposed to just movement choices, than the model with no territorial interactions.   To test whether the models are a good fit to the data, we used a Pearson's chi-squared test, treating each 10m by 10m square as a single data bin.  For this, we used the positional data rather than the smoothed data.

\section*{Results}

\subsection*{Framing existing models as coupled step selection functions}

\noindent{\bf Step selection and resource selection.} Step selection functions are simply single examples of equation (\ref{cssf_general}) with the collective term ${\mathcal C}_i({\bf x},{\bf y},{\mathcal P}_i^t)$ equal to $1$ \citep{fortinetal2005, foresteretal2009, vanakaetal2013, latombeetal2013}.  In other words we just consider one animal at a time, and how it interacts with its environment, without attempting to use the results to construct a mechanistic model of interacting animals.  Resource selection functions are similar, but the environment-independent movement term $\phi_i({\bf x}|{\bf y},\theta_0)$ is replaced with an availability function, which can take whichever form the user feels is appropriate for study, e.g. \citep{rhodesetal2005, boyceetal2002}.

\vspace{3mm}\noindent{\bf Individual based territory models.} The selection of studies by \citet{GPH1, GPH3} and \citet{PHG3,PHG2} modelled territorial interactions using moving agents on a square lattice.  The initial model from \citep{GPH1} has agents performing nearest neighbour random walks and depositing scent as they move.  The scent remains for a finite time $T_\ast$, the so-called {\it active scent time}, after which it is no longer considered as `active' by conspecifics.  Each animal's movement is restricted by the fact that it cannot move into an area that contains active scent of a neighbour.

This can be framed as a coupled step selection function where $\phi_i({\bf x}|{\bf y},\theta_0)=1/4$ if ${\bf x}$ is the lattice site either immediately above, below, to the right, or to the left of ${\bf y}$, and $\phi_i({\bf x}|{\bf y},\theta_0)=0$ otherwise.  Additionally, since this model does not include any environmental interactions, we set ${\mathcal W}_i({\bf x},{\bf y},{\mathcal E}) = 1$.  The term ${\mathcal P}_i^t({\bf x})$ represents the presence of scent at position ${\bf x}$ and time $t$, so that
\begin{align}
{\mathcal P}_i^t({\bf x})=\begin{cases}T_\ast & \text{any animal $j\neq i$ is at position ${\bf x}$ at time $t$,}
\\
\min\{{\mathcal P}_i^{t-\tau}({\bf x})-\tau,0\} & \text{otherwise.}
\end{cases}
\label{fox_population}
\end{align}
Then the collective interaction term is
\begin{align}
{\mathcal C}_i({\bf x},{\bf y},{\mathcal P}_i^t)=\begin{cases}1 & \text{if ${\mathcal P}_i^t({\bf x})=0$,}
\\
0 & \text{otherwise.}
\end{cases}
\label{fox_interaction}
\end{align}
The coupled step selection function formalism (equation \ref{cssf_general}) gives a natural way of incorporating environmental interactions into such territoriality models, an aspect of this approach hitherto lacking, as noted in \citep{GPRL}.

\vspace{3mm}\noindent{\bf Advection-diffusion territory models.} The type of territorial models described in \citep{moorcroftlewis} provide several other examples of coupled step selection functions.  We describe an individual-level model in a 1D interval $[0,1]$ that has as its continuum limit the original advection diffusion model of \citep{lewismurray1993}.  To do this, we first set
\begin{align}
\phi_i({ x}|{y})=\frac{\exp(-|x-y|/a)}{2a},
\label{phi_i_lewis_terr}
\end{align}
where $a$ is the average step length, and ${\mathcal W}_i({x},{y},{\mathcal E}) = 1$.  This means that the intrinsic movement of each agent (pack of wolves) is a random walk with no correlation, and we are ignoring the effects of the environment on movement.

There are two agents in the model, so $i\in \{0,1\}$.  The collective action is mediated by scent deposition so that ${\mathcal P}_i^t({x})$ represents the scent mark density of pack $1-i$.  Marking by individual $i$ occurs at a rate $l+m{\mathcal P}_{1-i}^t({x})$, where $m$ is typically a monotonic increasing function, representing the tendency of wolves to mark more heavily when conspecific marks are present.  ${\mathcal P}_{i}^t({x})$ is governed by the following equation
\begin{align}
{\mathcal P}_i^{t}({x})=\underbrace{(1-\mu\tau){\mathcal P}_i^{t-\tau}({x})}_{\text{scent decay}}+\underbrace{\delta(x_{i-1}-x)[l+m{\mathcal P}_{1-i}^t({x})]\tau}_{\text{scent deposition}}
\label{wolf_population}
\end{align}
where $x_i$ is the position of agent $i$ at time $t-\tau$ and $\mu$ is the scent decay rate.

\renewcommand{\arraystretch}{0.4}

Packs have a tendency to move back towards their home range centre on encountering foreign scent.  Assuming that the home range center of pack 0 is to the left of the study area and pack 1 to the right, the collective interaction term is given by
\begin{align}
{\mathcal C}_0({x},{y},{\mathcal P}_0^t)=&
I(x>y)\tau [D/a-v{\mathcal P}_0^{t-\tau}(x)]+I(x\leq y)\tau[D/a+v{\mathcal P}_0^{t-\tau}(x)]
\\
{\mathcal C}_1({x},{y},{\mathcal P}_1^t)=&
\underbrace{I(x>y)\tau [D/a+C{\mathcal P}_1^{t-\tau}(x)]}_{\text{\begin{tabular}{c}right-ward drift due \\ to conspecific scent\end{tabular}}}+ \underbrace{I(x\leq y)\tau[D/a-C{\mathcal P}_1^{t-\tau}(x)]}_{\text{\begin{tabular}{c}left-ward drift due \\ to conspecific scent\end{tabular}}}
\end{align}
where $D$ and $C$ are parameters, which can be determined by model fitting, and $I(X)$ is an indicator function that is equal to 1 if $X$ is true and 0 otherwise.

Now we move from an individual description to positional probability density functions.  Let $u(x,t)$ (resp. $v(x,t)$) be the probability distribution of pack 0 (resp. pack 1).  For notational convenience, we rename the scent levels of packs 0 and 1 to $p(x,t)$ and $q(x,t)$ respectively.
Then standard theory, e.g. \citep[chapter 2]{moorcroftlewis}, means that the limit as $\tau\rightarrow 0$, $a \rightarrow 0$ of $u(x,t)$ is governed by the following advection-diffusion equation
\begin{align}
\frac{\partial u}{\partial t}=\underbrace{\frac{\partial^2}{\partial x^2}[d_u(x,t)u(x,t)]}_{\text{\begin{tabular}{c}random \\ movement\end{tabular}}}-\underbrace{\frac{\partial}{\partial x}[c_u(x,t)u(x,t)]}_{\text{\begin{tabular}{c}directed \\ motion\end{tabular}}},
\label{lewis_terr_me}
\end{align}
\renewcommand{\arraystretch}{1}
where the advection and diffusion functions [$c_u(x,t)$ and $d_u(x,t)$ respectively] are the following limits
\begin{align}
c_u(x,t)&=\lim_{\tau\rightarrow 0}\frac{1}{\tau}\int_{-\infty}^{\infty}(y-x)\phi_0({ x}|{y}){\mathcal C}_0({x},{y},q){\rm d}y, \nonumber \\
d_u(x,t)&=\lim_{\tau\rightarrow 0}\frac{1}{\tau}\int_{-\infty}^{\infty}(y-x)^2\phi_0({ x}|{y}){\mathcal C}_0({x},{y},p){\rm d}y.
\label{adv_diff}
\end{align}
This theory is built by constructing the master equation for $u$.  Implicit in the construction is the so-called `mean-field' approximation, which assumes that the covariance between the scent mark density and the position of the pack is (approximately) zero.  A direct calculation shows that $c_u(x,t)=Cq(x,t)$ and $d_u(x,t)=D$.  The equation for $v(x,t)$ is analogous, but with $\phi_0$, ${\mathcal C}_0$, $c_u$, $d_u$, and $q$ replaced by $\phi_1$, ${\mathcal C}_1$, $c_v$, $d_v$, and $p$ respectively.  Therefore $c_v(x,t)=-Cp(x,t)$ and $d_v(x,t)=D$.

\renewcommand{\arraystretch}{0.4}

The advection diffusion equations for this system of coupled step selection functions are then
\begin{align}
\frac{\partial u}{\partial t}=&D\frac{\partial^2u}{\partial x^2}-C\frac{\partial}{\partial x}[qu], \nonumber \\
\frac{\partial v}{\partial t}=&\underbrace{D\frac{\partial^2v}{\partial x^2}}_{\text{\begin{tabular}{c}random \\ movement\end{tabular}}}+\underbrace{C\frac{\partial}{\partial x}[pv]}_{\text{\begin{tabular}{c}directed \\ motion\end{tabular}}}.
\label{fin_adv_diff}
\end{align}
Furthermore, the continuous-time limits of the scent marking equations (\ref{wolf_population}) are as follows \citep[chapter 3]{moorcroftthesis}
\begin{align}
\frac{\partial p}{\partial t}=&u(l+mq)-\mu p, \nonumber \\
\frac{\partial q}{\partial t}=&\underbrace{v(l+mp)}_{\text{\begin{tabular}{c}scent \\deposition\end{tabular}}}-\underbrace{\mu q}_{\text{\begin{tabular}{c}scent \\decay\end{tabular}}}.
\label{scent_ode}
\end{align}
Equations (\ref{fin_adv_diff}) and (\ref{scent_ode}) form the system studied in \citep{lewismurray1993}.  This process can be generalized to derive advection diffusion equations describing territorial pattern formation in two dimensions \citep{moorcroftlewis}.

\renewcommand{\arraystretch}{1}


\vspace{3mm}\noindent{\bf Alignment-and-attraction models.} Equation (\ref{cssf_general}) also reduces to a variety of collective motion models other than territorial ones, including trail-following ants \citep{deneubourgetal1989} and collective patterns in animal populations through alignment and attraction \citep{couzinetal2002, eftimieetal2007}.  Here we address one of these modelling frameworks \citep{couzinetal2002} with the others left to the Supplementary Information.


To write the model from \citep{couzinetal2002} as a CSSF, we first notice that each animal, $i$, has a fixed speed, $s_i$.  Therefore we set $\phi_i({\bf x}|{\bf y},\theta_0)=\delta_D(|{\bf x}-{\bf y}|-s_i\tau)$, where $\delta_D$ is the Dirac delta function.  ${\mathcal W}_i({\bf x},{\bf y},{\mathcal E}) = 1$ since there are no environmental interactions in the model from \citep{couzinetal2002}.  All the other animals in the population can influence animal $i$'s subsequent movement, so $${\mathcal P}_i^t=({\bf y}_1,\dots,{\bf y}_{i-1},{\bf y}_{i+1},\dots,{\bf y}_n,\theta_1,\dots,\theta_{i-1},\theta_{i+1},\dots,\theta_n),$$ where ${\bf y}_j$ is the position of animal $j$ at time $t$, having arrived there on a bearing of $\theta_j$.

The model incorporates attraction, alignment and repulsion.  Repulsion occurs if there are other animals within distance of $r_r$ from animal $i$, to ensure that animals do not collide.  If there is no repulsion then animal $i$ will align with any others that are greater than a distance of $r_r$, but less than a distance of $r_o$, from $i$.  They will also be attracted to animals $j$ such that $r_o \leq |{\bf y}_j-{\bf y}_i|\leq r_a$ (see \citep{couzinetal2002} for details).

\renewcommand{\arraystretch}{0.4}

To aid in writing the interaction term, we let $\theta_r(P_i^t)$ be the repulsion angle, which is the bearing given by the vector
\begin{align}
{\bf v}_r=-\sum_{j\neq i}\underbrace{\frac{{\bf y}_j-{\bf y}_i}{|{\bf y}_j-{\bf y}_i|}}_{\text{\begin{tabular}{c}vector \\ from $i$ to $j$\end{tabular}}}\underbrace{I(|{\bf y}_j-{\bf y}_i|<r_r)}_{\text{\begin{tabular}{c}repulsion \\ radius\end{tabular}}}.
\label{avoid_vector}
\end{align}
We also define an alignment and attraction angle, $\theta_a(P_i^t)$, which is the bearing given by the direction of
\begin{align}
{\bf v}_a=&\underbrace{\sum_{j\neq i}\frac{{\bf y}_j-{\bf y}_i}{|{\bf y}_j-{\bf y}_i|}I(r_o\leq|{\bf y}_j-{\bf y}_i|\leq r_a)}_{\text{attraction}}+ \nonumber \\
&\underbrace{\sum_{j\neq i}\left( \begin{array}{c}
\cos(\theta_i) \\
\sin(\theta_i) \end{array} \right)I(r_r\leq|{\bf y}_j-{\bf y}_i|< r_o)}_{\text{alignment}}.
\label{attr_align_vector}
\end{align}
\renewcommand{\arraystretch}{1}
The interaction term from \citep{couzinetal2002}, section `Behavioural rules: description', is then
\begin{align}
{\mathcal C}_i({\bf x},{\bf y},{\mathcal P}_i^t)=\begin{cases}\mbox{SG}(\theta-\theta_r) & \text{if there is a $j\neq i$ such that $|{\bf y}_j-{\bf y}_i|<r_r$,}
\\
\mbox{SG}(\theta-\theta_a) & \text{if there is a $j\neq i$ such that $|{\bf y}_j-{\bf y}_i|\leq r_a$}
\\                         & \text{but no $k\neq i$ such that $|{\bf y}_k-{\bf y}_i|<r_r$,}
\\
\mbox{SG}(\theta-\theta_0) & \text{otherwise,}
\end{cases}
\label{couzin_interact}
\end{align}
where $\mbox{SG}(\psi)$ is a spherical Gaussian.

\subsection*{The example of Amazonian bird flocks}

When we apply our technique to data on Amazonian birds, there is no significant improvement in fit ($p=0.60$) if we model birds as having a tendency not to go into areas from where they have heard conspecific bird calls in the past (hypothesis 1 from the Methods section).  However, when flocks are modelled as being allowed to move into neighbouring territories, but then having a tendency to retreat in the direction of the gathering point (hypothesis 2), we observe a significant improvement in fit ($p=0.022$).  
If we assume that the territorial cues have a finite lifetime (hypothesis 3), the maximum likelihood estimator for $T_\ast$ is larger than the length of the time series data, suggesting that birds are able to remember these cues for a very long time after they have been made.

\begin{figure}[ht!]
\centering
\includegraphics[width=150mm]{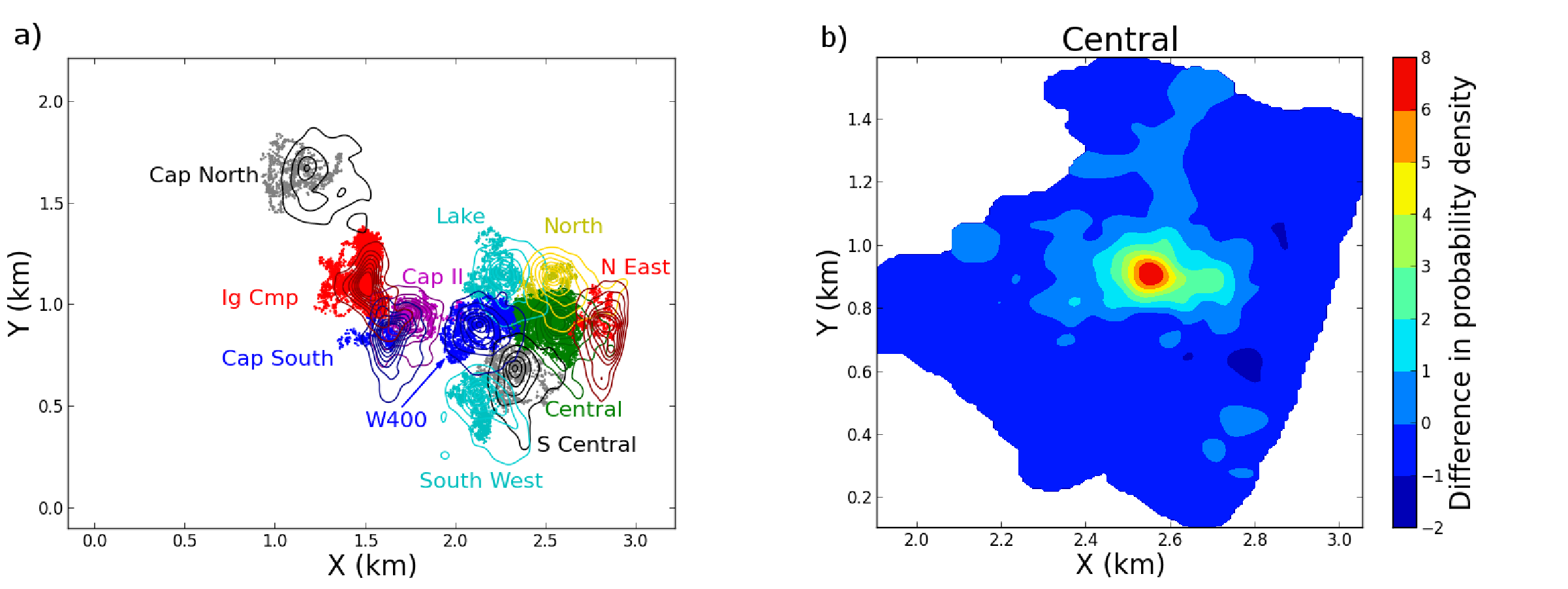}
\caption{{\bf Space use predictions of bird flocks using a coupled step selection function.}  In panel (a), the dots represent recorded positions of bird flocks, whereas the contours detail the space use distributions that arise from a territorial- and environmental-interaction model that best fits the movement data (see Methods for details).  The colors of the contours for each flock correspond to those of both the positional data points and the text giving the flock names.  Panel (b) shows the predicted position distributions for the Central flock with territorial interactions minus those without such interactions. Note that no fitting was performed between the model spatial distribution and the bird positions.  Instead, the distributions simply emerge from the model's underlying movement and interaction processes.}
\label{flock_space_use}
\end{figure}

To demonstrate the space use patterns that arise from these results, we constructed simulations using the gathering point attraction model, used to test hypothesis 2, with the best fit parameters of $T_\ast=\infty$ and $\kappa=0.0597$ (Fig. \ref{flock_space_use}).  For 9 of the 11 flocks, the resulting Kernel Density Estimator (KDE) distributions are closer to those of the data than the KDE distributions without territorial interactions (see Table \ref{kl_table}).  Furthermore, the resulting difference in Akaike Information Criteria ($\Delta$AIC) between the two models is $\Delta$AIC$=4.07$, giving reasonable evidence to suggest that the model including territorial interactions is better at predicting space use patterns than that without.  This is demonstrated pictorially in Fig. \ref{flock_space_use}b, which shows that the model including territorial interactions is more highly peaked at the center and includes a lower density of outliers.

Of the two flocks that are not well-modelled by incorporating territorial interactions, for Cap North we have no data on adjacent flocks (Fig. \ref{flock_space_use}a) so the inability of the model to detect territorial interactions is unsurprising.  Cap II, on the other hand, is located in the most degraded area of all flocks in the study. Subsequent observations of the study area suggest that it did not persist over time, as key species either abandoned the area or died.  Therefore the territory could well be in the process of moving or degrading during the study period, mechanisms that are likely to be key drivers in shaping the space use, but which are absent from our current model.

For all of the flocks except Cap II, there was insufficient evidence to suggest that the data did not come from the model distribution that included territorial interactions ($p<0.0001$ for Cap II, $p>0.999$ for the others).  The same test with the model that excluded territorial interactions suggested that there was only sufficient evidence to reject the hypothesis that the data came from the model for Cap II and Central ($p<0.0001$ for Cap II and Central, $p>0.999$ for the others).  Therefore we have significantly improved the absolute fit of the Central data by including territorial interactions.  Central is the only flock for which we have data on all surrounding flocks so it is precisely the flock for which one would most expect to see improvement of absolute fit.



\section*{Discussion}

We have constructed a general model for the effects on movement of both animal-habitat and between-animal interactions.  We have demonstrated how the model encompasses, as special cases, a variety of disparate collective motion models as well as resource and step selection functions.  By fitting a version of our model to data on bird flock locations, we have shown how it can be used to determine and quantify the nature of territorial interactions, as well as modelling simultaneously the effects of both conspecifics and the environment on movement processes.  Since we framed the system as a one-step Markovian model of both the animals and their environment, our framework allows for relatively simple calibration of models, which makes the process computationally fast.  This contrasts with methods that fit the movement path as a whole, such as state-space models, which can be difficult to fit \citep{pattersonetal2008}.


Though we have focused on territorial modelling, so not given an exhaustive demonstration of how our framework might be reducible to all collective behaviour models in the literature, we display a variety of different examples, encompassing both direct and mediated interactions, both conspecific attraction and avoidance processes.  These demonstrate the possible wide applicability of our approach, and potential to frame many more models as coupled step selection functions.  Encompassing competing models of collective behaviour under this unifying framework will make future comparisons easier, aided by the methods given here for fitting coupled step selection functions to data.  Furthermore, it will enables transference of techniques and results between the hitherto disparate fields of collective motion, resource selection and mechanistic territorial modelling.  To give one example, research into ungulate behaviour often looks at the effects of the environment on movement but ignores herding interactions (e.g. \citep{fortinetal2005}), or looks at herding behaviour but ignores the resource aspect (e.g. \citep{gueronlevin1993}).  Our framework links these two ideas so will help future researchers build and validate models that account for both.


By applying our model to movement patterns of bird flocks, we were able to test hypotheses about the mechanisms behind the interaction processes.  Previous studies of mechanisms underlying territorial patterns in populations of scent-marking animals postulated that they will avoid areas that have recently been claimed by others as their territory \citep{PHG3}.  Here we have shown that the territorial interaction mechanism in bird flocks is quite different.  There is no evidence to suggest that they tend to avoid places that have previously been claimed as other flocks' territories.  However, after visiting the outskirts of neighbouring territories, they will change their movement processes to include a tendency to retreat back inside their territory.  These visitations explain the observed slightly overlapping utilisation distributions in the birds' spatial patterns (see Fig. \ref{flock_space_use}).


Our framework can also be used to build predictive, mechanistic models showing how utilisation distributions arise from the underlying movement and interaction processes.  To demonstrate this, we used stochastic simulations of the best fit system of coupled step selection functions for the bird data. Recently, step selection functions have been used to construct deterministic master equation \citep{PMSL1} and partial differential equation models \citep{moorcroftbarnett2008}, from which the resulting spatial distributions can be analysed using well-studied mathematical tools, e.g. \citep{moorcroftlewis}.  Whilst the coupling term in our framework makes such analysis significantly more complicated than for ordinary step selection functions, deterministic mathematical formulations would ultimately enable concrete conclusions to be reached without the need for extensive, time-consuming computer simulations.  We therefore hope, in future work, to begin a program of analysing coupled step selection models mathematically.

Though mechanistic models have previously been proposed to explain space use patterns by examining both movement, territorial interactions and environmental features  \citep{moorcroftetal2006}, those models fit the emergent space use distribution to relocation data, whereas our model is directly fitted to the movement trajectory itself, enabling the space use distribution to arise with no additional fitting.  The advantage of this is twofold.  First, there is no need to throw away data in order to make sure each data point is an independent sample of the spatial distribution from the others (see \citep{moorcroftlewis} for details of, and rationale behind, this procedure).  Therefore we can use the complete movement trajectory, containing much more information.

Second, fitting the model to the underlying movement choices ensures that the parameter values used to construct the model arise from the movement and interaction processes rather than the emergent patterns.  This means that we can assess to what extent these processes predict space use, and where they fail.  For example, in the data studied here, the space use of two flocks (Cap II and Cap North) were not predicted by the territorial interaction model as well as by the no-interaction model, unlike the other nine flocks (Table \ref{kl_table}).  Therefore we can postulate hypotheses about what other processes may be required to predict space use in these instances.  On the other hand, fitting directly to the space use distribution implicitly assumes that the mechanistic model describes well all aspects of movement that give rise to the spatial patterns.  Consequently, this procedure may cause inaccurate inferences to be made about the parameter values of the underlying processes.  In other words, our approach is more cautious, therefore less likely to lead to incorrect results and more likely to reveal the extent to which certain processes fail to predict accurately the spatial patterns.

As an alternative to mathematical models of space use, simulations of individual based models have also been used to attempt to understand animal movement decisions and emergent spatial patterns \citep{hartigetal2011}.  Typically, they take a pattern-oriented approach \citep{grimmetal1996, grimmetal2005}, beginning by including as many aspects of the animal's movement and interaction processes as are believed to cause the observed patterns.  If the empirical patterns, also called {\it summary statistics}, are observed in the model output then the model is simplified to try to understand exactly which of the processes are causing the patterns to emerge.  The aim of this approach is to find models that replicate as many of the summary statistics observed in the data as possible, with as few model parameters.

Our approach, on the other hand, is {\it process-based} in nature \citep{evansetal2013}, seeking to build an individual based mechanistic model by testing hypotheses about the underlying processes one at a time.  The key difference is that we test the model parameters against the data for validity on the same level of description at which the model is constructed.  The pattern-oriented approach tests the model parameters at a different level of description: that of the summary statistics.  However, this is not sufficient for making inferences about the parameter values put into the model.  Though analysis of a mechanistic model, individual based or analytic, shows that process $A$ implies pattern $B$, showing that pattern $B$ replicates the data does not imply that the underlying mechanism is actually process $A$.  Therefore it is not possible, purely using a pattern-oriented approach, to make solidly-grounded inferences about the nature of the mechanisms that have gone into construction of the model.  In our approach, we circumvent this issue by testing and parameterising the model's mechanisms on the level of description at which they are constructed, then observing the patterns as an emergent feature of the model, which can in turn be compared with the patterns from data.

Recent developments in the collective behaviour literature provide many good examples of process-based modelling and model parameterisation \citep{nagyetal2010, herbertreadetal2011,katzetal2011,gautraisetal2012,strandburgpeshkinetal2013,perezescuderopolavieja2011,mannetal2014, pettitetal2013}.  However, very few examine the emergent features of these data-parameterised models and test whether they accurately replicate the population level patterns seen in the data, as we do here.  That said, there are exceptions, e.g. \citep{pettitetal2013,bodeetal2010, bodeetal2011}, and these models could, in principle, be used in conjunction with theoretical mechanistic models of pattern formation, such as \citep{eftimieetal2007,goldstoneetal2005}, to provide a full story.  If they were to be framed under a single overarching methodological framework, such as the coupled step selection functions proposed here, then this would aid sort of the unification of process-based model construction and theoretical process-to-pattern analysis that has recently been sought after \citep{sumpteretal2012}.

Though our model was significantly better at predicting space use than the model free of territorial interactions, it is clear from Fig. \ref{flock_space_use}a that our model does not capture all aspects of the birds' spatial patterns.  However, the strength of our approach is that we can readily add further behavioural features one at a time, testing the efficacy of each using the techniques detailed here.  For example, the birds are known to have direct territorial conflicts, which affect where they move in subsequent days and weeks.  Also, the movement is driven by intra-flock interactions, with one particular species, the Cinerous Antshrike ({\it Thamnomanes caesius}), playing the main role in maintaining cohesiveness.  By using our techniques to test the effect of such behavioural phenomena on movement and space use, we can move towards building truly accurate, predictive models linking movement processes, conspecific interactions and collective behaviour, to the emergent space use distributions.

\section*{Acknowledgments}
This study was partly funded by NSERC Discovery and Accelerator grants (MAL, JRP).  MAL also gratefully acknowledges a Canada Research Chair and a Killam Research Fellowship.  KM would like to acknowledge the Biological Dynamics of Forest Fragments Project (BDFFP) staff for providing logistic support; J. Lopes, E.L. Retroz, P. Hendrigo, A. C. Vilela, A. Nunes, B. Souza, M. Campos for field assistance; M. Cohn-Haft for valuable discussions. Funding for the research was provided by US National Science Foundation grant LTREB 0545491 awarded to Phil Stouffer, which helped fund KM's work, and by the AOU 2010 research award to KM. This article represents publication no. 643 in the BDFFP Technical Series. This is contribution no. 33 in the Amazonian Ornithology Technical Series of the INPA Zoological Collections Program. This manuscript was approved for publication by the Director of the Louisiana Agricultural Experiment Station as manuscript 2014-241-15478. LIDAR images for canopy height models and digital elevation models were provided by Scott Saleska (University of Arizona) and Michael Lefsky (Colorado State University).  We are grateful to Phil Stouffer, Greg Breed and Andrew Bateman for helpful discussions and suggestions, as well as several others who examined previous versions of our manuscript, including four anonymous referees.




\newpage

\newpage


\newpage


\renewcommand{\arraystretch}{1}

\newpage
\begin{table}
\caption{Various models from the literature that can be formulated as coupled step selection functions.  Interaction models are classified as one or more of animal-environment (E), between-animal direct interactions (BD), between-animal mediated interactions (BM), alignment-and-attraction models (AA), conspecific avoidance models (CA). }
\begin{tabular}{@{\vrule height 10.5pt depth4pt  width0pt}l|c|c|c}
  Model & Reference & Interaction type \\
\hline
  Resource selection & \citet{boyceetal2002} & E \\
  Step selection & \citet{fortinetal2005} & E \\
  Individual-based collective behaviour & \citet{couzinetal2002} & BD, AA \\
  Differential equation collective behaviour & \citet{eftimieetal2007} & BD, AA \\
  Army ant foraging & \citet{deneubourgetal1989} & BM, AA \\
  Individual-based territory formation & \citet{GPH1} & BM, CA \\
  Differential equation territory formation & \citet{moorcroftlewis} & BM, CA \\
\end{tabular}
\label{cssf_lit_table}
\end{table}

\newpage
\begin{table}
\caption{Fitting models both with and without territorial interactions to data on bird flock movement.  For each flock, the Kullbeck-Leibler (K-L) distance between the data's Kernel Density Estimator (KDE) distribution and the model's KDE distribution is given.  For all but two of the flocks, the model that includes territorial interactions performs best, shown by a positive difference in column 4.}
\begin{tabular}{@{\vrule height 10.5pt depth4pt  width0pt}lccc}
  Flock & K-L with interactions & K-L no interactions & Difference \\
\hline
Central &	0.868 &	1.236 &	0.367 \\
North &	1.018	& 1.442	& 0.424 \\
South Central & 0.673 & 0.826 & 0.152 \\
South West & 1.020 & 1.317 & 0.297 \\
Lake & 0.902 & 1.063 & 0.161 \\
W400 & 0.737 & 0.989 & 0.252 \\
Cap II & 3.527	& 3.377	& -0.150 \\
Cap South &	1.192 & 1.465 & 0.305 \\
Ig-Cmp & 0.779	& 1.013	& 0.234 \\
Cap North &	1.125	& 1.048	& -0.077 \\
North-East & 0.967	& 1.038	& 0.071 \\
\end{tabular}
\label{kl_table}
\end{table}




\end{document}